\documentclass[aps,prl,reprint,amsmath,amssymb]{revtex4-2}

\usepackage{bm}
\usepackage{graphicx}
\usepackage{amsthm}
\usepackage{xcolor}
\usepackage[colorlinks=true,linkcolor=red!80!black,citecolor=blue!80!black,urlcolor=blue!80!black]{hyperref}

\newtheorem{theorem}{Theorem}
\newtheorem{proposition}{Proposition}
\newtheorem{lemma}{Lemma}
\newtheorem{corollary}{Corollary}
\newtheorem{definition}{Definition}

\newcommand{\ip}[2]{\langle #1,#2\rangle}
\newcommand{\C}{\mathrm{C}}
\newcommand{\AQ}{\mathrm{AQ}}
\newcommand{\Id}{\openone}

\begin{document}

\title{Almost-Quantum Correlations Violate Information Causality}

\author{Anandamay Das Bhowmik}
\affiliation{S. N. Bose National Centre for Basic Sciences, Block JD, Sector III, Salt Lake, Kolkata 700106, India}

\author{Manik Banik}
\affiliation{S. N. Bose National Centre for Basic Sciences, Block JD, Sector III, Salt Lake, Kolkata 700106, India}

\begin{abstract}
\noindent Information causality (IC) posits that communicating $m$ classical bits cannot give the receiver access to more than $m$ bits of information about the sender’s independent data. Although IC holds in quantum theory and excludes a broad class of supra-quantum correlations, its validity for the almost-quantum set has remained open. This semidefinite relaxation of the quantum set satisfies nontrivial communication complexity, no advantage for non-local computation, macroscopic locality, local orthogonality, and numerical evidence has suggested that it may also satisfy IC. Here we show that almost-quantum correlations violate IC, and do so without bound. Unlike earlier violations of IC, which rely on biased guesses, ours is driven by zero-error certificates: each is rare but never wrong, carries information linear in its probability, and vouches for many of the sender’s bits at once. This establishes a sharp dichotomy: either IC is a valid physical principle that excludes these correlations, or any nonsignalling theory realizing them admits no mutual information obeying consistency, data processing and the chain rule. In the latter case, we discuss why the chain rule is the natural property to relinquish, and how this possibility relates to histories-based formulations and to locality defined relative to the state rather than presupposed in a fixed spacetime.
\end{abstract}

\maketitle
\noindent
Bell's theorem~\cite{Bell1964} marks one of the sharpest departures of quantum theory from classical physics~\cite{Bell1966,Mermin1993,Brunner2014}. Addressing the question raised by Einstein, Podolsky and Rosen of whether the quantum-mechanical description of physical reality is complete~\cite{EPR1935}, Bell derived an inequality that the statistics of local measurements on separated systems must obey whenever they admit a local-causal explanation, and showed that quantum mechanics predicts its violation~\cite{Bell1964,CHSH1969}. Such violations have since been observed in several milestone experiments~\cite{FreedmanClauser1972,Aspect1981,Aspect1982a,Aspect1982b,Weihs1998,Hensen2015,Giustina2015,Shalm2015}. Beyond its foundational significance, Bell nonlocality is a resource for device-independent information processing, in which conclusions are drawn from the observed statistics alone, without trusting the internal workings of the devices~\cite{Scarani2012,Brunner2014}. Examples include quantum key distribution based on Bell tests and its device-independent versions, randomness certification and amplification, self-testing of quantum states and measurements, and device-independent witnesses of entanglement and of Hilbert-space dimension along with other applications~\cite{Ekert1991,MayersYao1998,Acin2007,VaziraniVidick2014,ArnonFriedman2018,Nadlinger2022,Zhang2022,Pironio2010,ColbeckRenner2012,Gallego2013,AcinMasanes2016,MayersYao2004,SupicBowles2020,Rai2021,Rai2022,Bancal2011,Brunner2008,Mukherjee2015,Cubitt2010,Alimuddin2023,Brunner2013,Roy2016,Banik2019}.

Quantum nonlocality is nevertheless restricted as compared to what relativistic causality alone would permit. For the Clauser--Horne--Shimony--Holt (CHSH) expression~\cite{CHSH1969}, local hidden-variable models reach at most $2$ and quantum theory at most $2\sqrt2$, Tsirelson's bound~\cite{Tsirelson1980}, whereas nonsignalling Popescu--Rohrlich (PR) boxes attain the algebraic maximum $4$~\cite{PR1994} (see also \cite{Popescu2014}). Explaining this gap, and ideally singling out the set of quantum correlations from physical or information-theoretic principles, is a long-standing programme. Proposed principles include nontrivial communication complexity~\cite{vanDam2013,Brassard2006,BrunnerSkrzypczyk2009}, no advantage for nonlocal computation~\cite{Linden2007}, information causality (IC)~\cite{Pawlowski2009}, macroscopic locality~\cite{NavascuesWunderlich2010} and local orthogonality~\cite{Fritz2013}. Each excludes a large class of supra-quantum correlations, but none is known to single out the quantum set. IC generalizes the no-signalling principle, which it contains as the case of zero communication: $m$ bits of communication cannot give the receiver access to more than $m$ bits of information about the sender's independent data. IC holds in quantum theory, recovers Tsirelson's bound~\cite{Pawlowski2009} and part of the boundary of the quantum set~\cite{Allcock2009}, and has been developed in several directions~\cite{Safi2011,Miklin2021,Patra2023,MiklinJainGachechiladze2026}.

Almost-quantum (AQ) correlations~\cite{Navascues2015} are the most demanding test case for this programme. They form the ``$1+AB$'' level of the Navascu\'es--Pironio--Ac\'in (NPA) hierarchy~\cite{NPA2007,NPA2008}, a set that strictly contains the quantum set yet is characterized by a single semidefinite program. They also arise from strongly positive decoherence functionals in Sorkin's quantum measure theory~\cite{Sorkin1994} when later measurements may depend on earlier outcomes~\cite{DowkerHensonWallden2014}, and from a principle of macroscopic noncontextuality~\cite{HensonSainz2015}. AQ correlations satisfy nontrivial communication complexity, no advantage for nonlocal computation, macroscopic locality and local orthogonality~\cite{Navascues2015}. For IC the question has remained open. IC is violated by the first level of the NPA hierarchy, which contains the AQ set~\cite{Cavalcanti2010}, whereas numerical evidence suggested that the AQ set satisfies it~\cite{Navascues2015}.

Here we show that AQ correlations violate IC, and do so without bound. A single use of one explicit AQ box and one classical bit suffice; Alice's data are independent fair bits, every run produces a guess, and nothing is postselected. The box, also studied in a companion paper~\cite{GaziCompanion} to separate AQ from commuting-quantum correlations in zero-error communication, provides certificates of equality: whatever the number $N$ of settings, its outcomes certify with probability $1/17$ that both parties chose the same setting, and they never do so when the settings differ. Quantum boxes can certify equality in this way with probability at most $1/N$~\cite{GaziCompanion}. Such certificates are exactly what IC cannot tolerate: an answer that is rarely given but always correct carries information linear in its probability, and a single certificate can vouch for many of Alice's bits at once.

\textit{Information causality.--} The principle of IC can be appreciated simply with the communication task of random access codes (RACs)~\cite{Wiesner1983,Ambainis2002,PawlowskiZukowski2010}. Alice receives $D$ independent, uniformly distributed bits $Z_1,\dots,Z_D$, and Bob receives an index $r\in\{1,\dots,D\}$. They share a nonsignalling resource that is independent of these inputs. Alice sends Bob an $m$-bit message, and Bob outputs a guess $G_r$ of $Z_r$. IC requires~\cite{Pawlowski2009}
\begin{equation}\label{eq:IC}
\mathcal I\equiv\sum_{r=1}^{D}I(Z_r:G_r)\le m,
\end{equation}
where $I(Z_r:G_r)$ is the Shannon mutual information in runs in which Bob's index is $r$. The sum runs over alternative questions, not over bits that Bob learns simultaneously. Pawłowski \textit{et al.}~\cite{Pawlowski2009} derived Eq.~\eqref{eq:IC} for any theory in which a mutual information $I(X:Y)$ can be defined for all pairs of systems with three properties: (i)~\emph{consistency}, it reduces to the Shannon mutual information when both systems are classical; (ii)~\emph{data processing}, local operations on either system cannot increase it; and (iii)~the \emph{chain rule}, $I(X:YW)=I(X:W)+I(X:Y|W)$. Data processing bounds each term of Eq.~\eqref{eq:IC} by the information that the message $M$ and Bob's share $\mathcal{B}$ of the resource carry about $Z_r$. The chain rule and the independence of the $Z_r$ bound the sum of these terms by the single quantity $I(Z_1\cdots Z_D:M\mathcal{B})$. Finally, because the resource is nonsignalling, $\mathcal{B}$ alone carries no information about Alice's data, and the $m$-bit classical message contributes at most $m$. The quantum mutual information has all three properties, so every protocol with quantum resources, including arbitrary shared entanglement, satisfies Eq.~\eqref{eq:IC}~\cite{Pawlowski2009}. A set of correlations violates IC if some protocol that uses correlations from the set violates Eq.~\eqref{eq:IC}. By the derivation above, no theory whose physical systems realize such correlations admits a mutual information with properties (i)--(iii).

\textit{Certificate boxes:} Consider a bipartite nonsignalling box $\bm p_s=\{p(a,b|x,y)\}$ with $N$ settings per party, $x,y\in\{1,\dots,N\}=[N]$, and binary outcomes $a,b\in\{0,1\}$, such that
\begin{equation}\label{eq:cert}
p(0,0|x,y)=s\,\delta_{xy}\qquad\text{for some } s>0 .
\end{equation}
The outcome pair $(0,0)$ then certifies with certainty that $x=y$, and it occurs with probability $s$ whenever $x=y$; we call such a box a certificate box and $s$ its certification rate. For commuting-operator correlations, and hence for all quantum correlations, a box with $N\ge3$ settings and $p(0,0|x,y)=0$ for $x\neq y$ obeys $\sum_xp(0,0|x,x)\le1$~\cite{GaziCompanion}, so every quantum certificate box has $s\le1/N$. Shared randomness attains this value: the parties share a uniformly random label and each outputs $0$ exactly when its setting equals that label. In contrast, the box $\bm p^\star$ of Table~\ref{tab:box} satisfies Eq.~\eqref{eq:cert} with $s=1/17$ for every $N$, and it is almost quantum.

\begin{definition}[Almost-quantum correlations~\cite{Navascues2015}]\label{def:aq}
A box $\bm p$ belongs to the AQ set $\C_\AQ$ if there exist a Hilbert space $\mathcal H$, a unit vector $\psi\in\mathcal H$, and projective measurements $\{A_{a|x}\}_a$ and $\{B_{b|y}\}_b$ on $\mathcal H$ such that, for all $a,b,x,y$, $A_{a|x}B_{b|y}\psi=B_{b|y}A_{a|x}\psi$ and $p(a,b|x,y)=\ip{\psi}{A_{a|x}B_{b|y}\psi}$.
\end{definition}

\noindent Quantum correlations, in the commuting-operator framework, require $[A_{a|x},B_{b|y}]=0$ as an operator identity; the AQ set requires commutation only on the state $\psi$, and it strictly contains the quantum set~\cite{Navascues2015}. That $\bm p^\star$ is almost quantum was shown in Ref.~\cite{GaziCompanion}; we discuss the short proof for completeness (see Lemma~\ref{lem:aq} in the Appendix). Paw{\l}owski \textit{et al.} used the RAC version of IC game to show that PR boxes violate IC~\cite{Pawlowski2009}. To show that the AQ box $\bm p^\star$ violates IC, we first formulate a zero-error certification task in the spirit of a RAC.

\begin{table}[t]
\caption{\label{tab:box}The AQ box $\bm p^\star(a,b|x,y)$. It depends only on whether the settings coincide, has marginals $p_A(0|x)=p_B(0|y)=3/17$, and satisfies Eq.~\eqref{eq:cert} with $s=1/17$ for every number of settings.}
\begin{ruledtabular}
\begin{tabular}{lcccc}
 & $p(0,0)$ & $p(0,1)$ & $p(1,0)$ & $p(1,1)$\\
\hline
$x=y$ & $1/17$ & $2/17$ & $2/17$ & $12/17$\\
$x\neq y$ & $0$ & $3/17$ & $3/17$ & $11/17$\\
\end{tabular}
\end{ruledtabular}
\end{table}

\begin{figure}[t]
\includegraphics[width=\columnwidth]{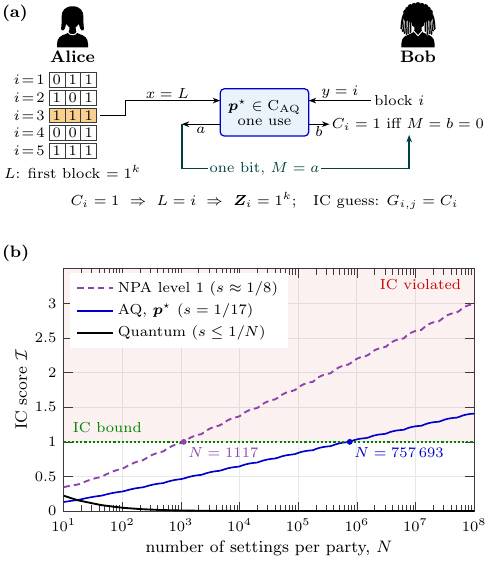}
\caption{\label{fig:main}(a) The block-certification protocol, drawn for $N=5$ blocks of $k=3$ bits. Alice inputs the index $L$ of the first block equal to $1^k$ and sends her outcome as one bit. Bob, asked about block $i$, inputs $y=i$ and certifies it, $C_i=1$, only if the message and his outcome both equal $0$. By Eq.~\eqref{eq:cert} this happens only if $L=i$, so a certificate is never false. In the IC game, Bob answers a question about any bit $Z_{i,j}$ of block $i$ with $G_{i,j}=C_i$. (b) The bound~\eqref{eq:score}, maximized over $k$, versus the number of settings per party $N$: for the AQ box $\bm p^\star$ ($s=1/17$, solid), for the level-1 NPA boxes of Ref.~\cite{GaziCompanion} with $s=N/[8(N-2)]$ (dashed), and for the largest certification rate compatible with quantum theory, $s=1/N$ (black). Dots mark the first violations of IC.}
\end{figure}

Random-access block certification (RABC): Alice receives a uniformly random string $\bm Z\in\{0,1\}^{kN}$ (equivalently, $kN$ independent fair bits), which we divide into $N$ blocks of $k$ bits, $\bm Z=(\bm Z_1,\dots,\bm Z_N)$ with $\bm Z_i=(Z_{i,1},\dots,Z_{i,k})\in\{0,1\}^k$ [Fig.~\ref{fig:main}(a)]. Bob receives a block index $i\in[N]$, independent of $\bm Z$. Alice sends Bob one bit $M\in\{0,1\}$, which may depend on $\bm Z$ but not on $i$, and Bob either certifies that block $i$ consists only of ones, $C_i=1$, or gives no certificate, $C_i=0$. A certificate must never be false:
\begin{equation}\label{eq:zeroerr}
\Pr\big(C_i=1,\ \bm Z_i\neq1^k\big)=0\qquad (i\in[N]),
\end{equation}
where $1^k$ denotes the all-ones string of length $k$. The figure of merit is the total certification probability $\mathcal W=\sum_{i=1}^{N}\Pr(C_i=1)$, where $\Pr(C_i=1)$ refers to runs in which Bob's index is $i$. As in a RAC, Bob is asked about one part of Alice's string, but, unlike in a RAC, he does not guess a bit; instead, he certifies, without error, that the whole block $\bm Z_i$ equals $1^k$, and he may always decline. Since $\Pr(\bm Z_i=1^k)=2^{-k}$, Eq.~\eqref{eq:zeroerr} gives $\Pr(C_i=1)\le2^{-k}$, and hence $\mathcal W\le N2^{-k}$ whatever the resources. With shared randomness alone, $\mathcal W\le2^{-k}$, and this value is attained if Alice reports whether a randomly chosen block equals $1^k$ (see Lemma~\ref{lem:classical} in the Appendix). 

Certificates obtained in the RABC game are valuable in the IC game for two reasons. A guess that is made with probability $t$ but is never wrong carries at least $t$ bits about the requested bit, that is, information linear in its probability. Moreover, a certificate for block $i$ reveals every bit of that block, so it answers correctly whichever of the $k$ bits $Z_{i,1},\dots,Z_{i,k}$ Bob is asked for. The following proposition makes both points precise.

\begin{proposition}\label{prop:info}
Take any zero-error RABC protocol. In the IC game with $D=kN$ bits, $Z_r\equiv Z_{i,j}$ for $r=(i-1)\,k+j$, let Alice send the same message as in the RABC protocol, and let Bob, asked for $Z_{i,j}$, run the protocol with block index $i$ and guess $G_{i,j}=C_i$. Then, $\mathcal I\ \ge\ k\,\mathcal W$; consequently, IC with $m=1$ implies $\mathcal W\le1/k$.
\end{proposition}

\begin{proof}
Alice's message does not depend on Bob's index, so the construction is a valid protocol for the IC game. Fix a question $(i,j)$ and let $t_i=\Pr(C_i=1)$. By Eq.~\eqref{eq:zeroerr}, a certificate requires $\bm Z_i=1^k$, so $t_i\le2^{-k}\le1/2$.

Let $Z$ be a fair bit and $G$ a binary guess with $\Pr(G=1,Z=0)=0$ and $\Pr(G=1)=t\le1/2$. The joint distribution is then $\Pr(Z=1,G=1)=t$, $\Pr(Z=1,G=0)=1/2-t$ and $\Pr(Z=0,G=0)=1/2$, yielding
\begin{equation}\label{eq:linear}
I(Z:G)=h(t)-\tfrac12h(2t)\ \ge\ t,
\end{equation}
with $h(\cdot)$ the binary entropy. For small $t$ the bound is nearly an equality: $h(t)-\tfrac12h(2t)=t+t^2/(2\ln2)+O(t^3)$, so the information carried by a certificate is linear in its probability. Since Bob's question is independent of Alice's data,  $Z_{i,j}$ is a fair bit. By Eq.~\eqref{eq:zeroerr}, the pair $(Z_{i,j},G_{i,j})=(Z_{i,j},C_i)$ satisfies the assumptions above with $t=t_i$. Hence $I(Z_{i,j}:G_{i,j})\ge t_i$. The guess $G_{i,j}=C_i$ does not depend on $j$, so each of the $k$ questions about block $i$ contributes at least $t_i$, and
\begin{align*}\mathcal I=\sum_{i=1}^{N}\sum_{j=1}^{k}I(Z_{i,j}:G_{i,j})\ \ge\ k\sum_{i=1}^{N}t_i=k\,\mathcal W .
\end{align*}
If IC holds, then $\mathcal I\le1$, and hence $\mathcal W\le1/k$.
\end{proof}

\textit{Block-certification protocol:} The bound $\mathcal W\le1/k$ of Proposition~\ref{prop:info} lies far above the classical value $2^{-k}$. A certificate box nevertheless exceeds it with the following block-certification protocol. Alice finds the first block that consists only of ones, $L=\min\{\,i:\ \bm Z_i=1^k\,\}$. If such a block exists, she inputs $x=L$ into the box, obtains $a$, and sends the one-bit message $M=a$. If no block consists only of ones, she sends $M=1$ and does not use the box. Bob, asked about block $i$, inputs $y=i$ and obtains $b$. He certifies, $C_i=1$, if $M=b=0$, and gives no certificate otherwise. Alice's message thus depends only on her string and her outcome, and Bob's setting only on his index.

\begin{theorem}\label{thm:main}
For a box obeying Eq.~\eqref{eq:cert} and every integer $k\ge1$, the block-certification protocol uses the box at most once and one bit of communication, never certifies falsely, and achieves
\begin{equation}\label{eq:Wcert}
\mathcal W=s\,\big[1-(1-2^{-k})^{N}\big].
\end{equation}
Consequently, in the IC game with $D=kN$ data bits,
\begin{equation}\label{eq:score}
\mathcal I\ \ge\ k\,s\,\big[1-(1-2^{-k})^{N}\big].
\end{equation}
\end{theorem}

\begin{proof}
Write $u=2^{-k}$ for the probability that a given block equals $1^k$. The blocks are independent, and the box is independent of Alice's string and of Bob's index. Bob certifies only if $M=0$ and $b=0$. Alice sends $M=0$ only if some block equals $1^k$, in which case she inputs $x=L$, and only if her outcome is $a=0$. Bob, asked about block $i$, inputs $y=i$. By Eq.~\eqref{eq:cert} the outcome pair $(a,b)=(0,0)$ has probability zero unless $x=y$. Hence $C_i=1$ implies $L=i$, so $\bm Z_i=1^k$, and Eq.~\eqref{eq:zeroerr} holds.

The event $L=i$ occurs exactly when $\bm Z_{i'}\neq1^k$ for all $i'<i$ and $\bm Z_i=1^k$, so
\begin{align*}
\Pr(L=i)=(1-u)^{i-1}u\qquad (i\in[N]).
\end{align*}
If $L=i$, the box is used with settings $x=y=i$ and returns $(a,b)=(0,0)$ with probability $p(0,0|i,i)=s$. If $L\neq i$, Bob does not certify: either no block equals $1^k$ and $M=1$, or $x=L\neq y=i$ and $(0,0)$ is impossible. Therefore
\begin{equation}\label{eq:ti}
t_i\equiv\Pr(C_i=1)=s\Pr(L=i)=s\,u\,(1-u)^{i-1}.
\end{equation}

Summing Eq.~\eqref{eq:ti} over the blocks, and using $u(1-u)^{i-1}=(1-u)^{i-1}-(1-u)^{i}$, we have
\begin{align*}
\mathcal W=\sum_{i=1}^{N}t_i=s\,\big[1-(1-u)^{N}\big],
\end{align*}
which is Eq.~\eqref{eq:Wcert}; here $1-(1-u)^N$ is the probability that at least one block equals $1^k$. Since the protocol is a zero-error RABC protocol, Proposition~\ref{prop:info} applies, with Bob guessing $G_{i,j}=C_i$ when asked for $Z_{i,j}$, and gives Eq.~\eqref{eq:score}.
\end{proof}

\begin{corollary}[Almost-quantum violates IC]\label{cor:violation}
The AQ box $\bm p^\star$ violates IC without bound. (i)~With $k=18$ and $N=757\,693$ settings, $\mathcal I\ge1.00000018>1$. (ii)~With $N=k2^k$, $\mathcal I\ge k(1-e^{-k})/17$, which grows without bound.
\end{corollary}

\begin{proof}
Part~(i) is a direct evaluation of Eq.~\eqref{eq:score} with $s=1/17$, using $D=13\,638\,474$ data bits. For part~(ii), write $u=2^{-k}$. For $0\le u<1$, $\ln(1-u)\le-u$. Hence $(1-u)^N=e^{N\ln(1-u)}\le e^{-Nu}$, and with $N=k2^k$, so that $Nu=k$, we obtain $(1-2^{-k})^{k2^k}\le e^{-k}$. Equation~\eqref{eq:score} then gives $\mathcal I\ge (k/17)\,(1-e^{-k})$.
\end{proof}

\noindent Three remarks are in order. First, quantum boxes never yield a violation in this protocol. A quantum certificate box has $s\le1/N$, and the union bound gives $1-(1-2^{-k})^N\le N2^{-k}$, so the right-hand side of Eq.~\eqref{eq:score} is at most $k\,s\,N2^{-k}\le k2^{-k}$. Since $k2^{-k}$ equals $1/2$ for $k=1,2$ and decreases thereafter, this bound never exceeds $1/2$, consistent with the validity of IC in quantum theory [Fig.~\ref{fig:main}(b)]. A violation requires certification rates far above this limit. It needs $ks>1$ together with $N\gtrsim2^k$, the scale on which the bracket in Eq.~\eqref{eq:score} approaches one; hence $s\gtrsim1/\log_2N$, whereas quantum boxes have $s\le1/N$. Correspondingly, at the threshold of Corollary~\ref{cor:violation}(i), $\bm p^\star$ gives $\sum_xp(0,0|x,x)=N/17\approx4.5\times10^4$, while every commuting-operator box with $p(0,0|x,y)=0$ for $x\neq y$ obeys $\sum_xp(0,0|x,x)\le1$~\cite{GaziCompanion}.

Second, the violation requires many settings. With $s=1/17$, the bound~\eqref{eq:score} exceeds one only for $k\ge18$, and hence only for $N\gtrsim2^{18}$; the precise threshold is $N=757\,693$ (Corollary~\ref{cor:violation}). Such scenarios lie far beyond the reach of numerical searches, consistent with the earlier numerical evidence~\cite{Navascues2015}. The protocol itself applies to every box obeying Eq.~\eqref{eq:cert}. The first level of the NPA hierarchy contains such boxes with $s=N/[8(N-2)]\to1/8$~\cite{GaziCompanion}; with $k=9$ they violate IC for every $N\ge1117$ [Fig.~\ref{fig:main}(b)], an independent confirmation that this level violates IC~\cite{Cavalcanti2010}.

Finally, what makes certificates so effective? A binary guess of a fair bit that is correct with probability $(1+E)/2$ carries $1-h\big((1+E)/2\big)=E^2/(2\ln2)+O(E^4)$ bits. IC therefore constrains squared biases, as in the condition $E_1^2+E_2^2\le1$ derived from it~\cite{Pawlowski2009,Allcock2009}. A certificate behaves differently: by Eq.~\eqref{eq:linear}, an answer given with probability $t$ but never wrong carries at least $t$ bits, linearly in $t$. In our protocol a certificate for block $i$ occurs with probability $t_i\le s2^{-k}$, but it vouches for all $k$ bits of that block, and the $N\gtrsim2^k$ blocks compensate for its rarity. The resulting score, $ks[1-(1-2^{-k})^N]$, approaches $ks$ and exceeds one once $ks>1$ and $N$ is large enough. Quantum boxes cannot exploit this mechanism, because their certification rate is at most $1/N$, whereas the AQ box keeps $s=1/17$ for every $N$.

\textit{Discussion.--} Our construction shows that almost-quantum correlations permit unbounded violations of information causality with one classical bit and a single use of a bipartite box. It thus provides explicit counterexamples to the compatibility of AQ correlations with IC suggested by numerical evidence~\cite{Navascues2015}. Earlier obstructions to AQ correlations invoked Specker's principle in contextuality scenarios~\cite{Gonda2018} and the no-restriction hypothesis for theories reproducing the AQ set~\cite{Sainz2018}. Our violation depends only on the bipartite input--output probabilities and the classical operations specified in the protocol.

Since IC is a \emph{proposed physical principle}, its violation does not by itself establish unphysicality. It nevertheless imposes a precise obstruction: no nonsignalling theory that realizes our family and permits the protocol can admit a mutual information with properties (i)--(iii). The violation alone does not identify which property fails. For the generalized mutual information of Wakakuwa and Murao, defined through optimal rates of classical communication over channels with general probabilistic outputs, consistency and data processing hold, and together with no-signalling the chain rule would imply IC~\cite{WakakuwaMurao2012}; within their framework, it is therefore the chain rule that must fail. Composite systems may remain well defined, but their operational information content cannot obey the usual decomposition into conditional contributions.

This obstruction has a concrete connection to generalized histories formulations. In bipartite Bell scenarios, the existence of a strongly positive joint decoherence functional reproducing the experimental probabilities, including branching measurement contexts, characterizes the AQ set~\cite{Sorkin1994,DowkerHensonWallden2014}. Our family therefore shows that strong positivity and consistency under outcome-dependent measurement choices can coexist with unbounded IC violations. These histories conditions alone do not recover the information constraints of quantum theory.

A broader possibility is that such violations signal the need for a revised operational account of subsystems and informational composition. Questions of this kind arise independently in causal-set approaches~\cite{Bombelli1987} and in frameworks with dynamical or indefinite causal structure~\cite{Hardy2007,OreshkovCostaBrukner2012}. These developments motivate examining alternatives to state-independent operator locality. AQ correlations provide a precise statistical relaxation: in their Hilbert-space characterization, distant projectors need commute only on the prepared state. Whether this relaxation acquires a physical interpretation in any such framework remains open. A physical completion might require a different information calculus, or additional dynamical and compositional requirements might exclude the offending correlations. Our result constrains these possibilities without selecting a particular spacetime mechanism.

Our construction uses many settings, leaving the economy of the violation as an important question. Multiple copies of smaller boxes, different encodings, or other AQ behaviours may yield more economical violations. The exact zeros $p(0,0|x,y)=0$ for $x\neq y$ make the certification mechanism transparent but are not essential to a finite violation. Determining useful noise thresholds, particularly as the number of settings grows, remains important. More broadly, rare conclusive outcomes contribute information at first order in their probability, whereas small symmetric prediction biases contribute only at second order in the bias. Certification tasks therefore provide a natural setting in which to uncover operational differences that average guessing performance alone can conceal.

\noindent\textit{Author contributions.} A.D.B. has pursued this problem over the past few years, in particular exploring how certainty, rather than bias, can lead to a violation of information causality. Both authors contributed to conceptualizing the idea, proving the results and preparing the manuscript.

\noindent\textbf{Note added.} The almost-quantum box used here is studied in detail in a companion paper, ``\emph{An Unbounded Separation of Almost-Quantum and Commuting-Operator Correlations in Zero-Error Communication}’’ by R.~Gazi, B.~Chatterjee, J.~Karmakar, R.~K.~Patra, M.~Banik and A.~Das~Bhowmik~\cite{GaziCompanion}. That paper separates almost-quantum from commuting-operator correlations in the zero-error certification of equality; its results do not concern information causality. The present work is self-contained in its main result.

\noindent{\bf Acknowledgement.} We thank R.~Gazi, B.~Chatterjee, J.~Karmakar and R.~K.~Patra, our coauthors on the companion paper~\cite{GaziCompanion}, for useful discussions. We also thank Sahil Gopalkrishna Naik, Mir Alimuddin, and Amit Mukherjee for useful discussions on the idea of certainty in IC scenario.  M.B. acknowledges financial support from the National Quantum Mission (NQM) of the Department of Science and Technology (DST), Government of India. 

\noindent{\bf Use of AI tools.} The problem was conceived and the principal results were obtained by the authors. The large language model Claude (Anthropic) was used to assist with editing the prose, checking the numerical thresholds and preparing the figure. 


%

\appendix

\section{Appendix A: Almost-quantumness of the box in Table~\ref{tab:box}}

\begin{lemma}\label{lem:aq}
For every $N\ge1$, the box $\bm p^\star$ of Table~\ref{tab:box}, with settings $x,y\in\{1,\dots,N\}$, belongs to $\C_\AQ$.
\end{lemma}

\begin{proof}
Let $\{\psi,e_1,e_2,f_x,g_x:x=1,\dots,N\}$ be an orthonormal basis of $\mathbb R^{2N+3}$, and define
\begin{align*}
&c_x=\tfrac{3}{17}\psi+\tfrac{\sqrt{15}}{34}\,e_2+\tfrac{\sqrt5}{2\sqrt{17}}\,g_x,\quad u_x=c_x+d_x,\\
&~~~~~d_x=\tfrac{\sqrt{51}}{34}\,e_1+\tfrac{1}{2\sqrt{17}}\,f_x,\quad v_x=c_x-d_x,\\
&~~~~~~~~~~w_x=\tfrac{1}{17}\psi-\tfrac{2\sqrt{15}}{85}\,e_2+\tfrac{2}{\sqrt{85}}\,g_x .
\end{align*}
These are the vectors $a_x$, $b_x$ and $w_x$ of Ref.~\cite{GaziCompanion}, with $u_x=a_x$ and $v_x=b_x$. Direct multiplication gives
$\ip{\psi}{u_x}=\|u_x\|^2=\ip{\psi}{v_x}=\|v_x\|^2=3/17$,
$\ip{\psi}{w_x}=\|w_x\|^2=\ip{u_x}{w_x}=\ip{v_x}{w_x}=\ip{u_x}{v_x}=1/17$, and
$\ip{u_x}{v_y}=\ip{u_x}{w_y}=\ip{v_x}{w_y}=0$ for $x\neq y$.
Let $A_{0|x}$ be the projector onto $\mathrm{span}\{u_x,w_x\}$ and $B_{0|y}$ the projector onto $\mathrm{span}\{v_y,w_y\}$, and set $A_{1|x}=\Id-A_{0|x}$ and $B_{1|y}=\Id-B_{0|y}$. By the inner products above, each of $\psi-u_x$, $v_x-w_x$ and $v_y$ ($y\neq x$) is orthogonal to both $u_x$ and $w_x$, so $A_{0|x}\psi=u_x$, $A_{0|x}v_x=w_x$ and $A_{0|x}v_y=0$; exchanging the roles of $u$ and $v$ gives $B_{0|y}\psi=v_y$, $B_{0|y}u_y=w_y$ and $B_{0|y}u_x=0$. Hence
$A_{0|x}B_{0|y}\psi=B_{0|y}A_{0|x}\psi=\delta_{xy}w_x$,
and the commutation on $\psi$ in Definition~\ref{def:aq} extends to all outcome pairs because $A_{1|x}=\Id-A_{0|x}$ and $B_{1|y}=\Id-B_{0|y}$. Finally, $p(0,0|x,y)=\delta_{xy}\ip{\psi}{w_x}=\delta_{xy}/17$ and $p_A(0|x)=\ip{\psi}{u_x}=p_B(0|y)=\ip{\psi}{v_y}=3/17$. For binary outcomes these values fix all entries of Table~\ref{tab:box}.
\end{proof}

\section{Appendix B: Classical value of RABC}

\begin{lemma}[Classical value]\label{lem:classical}
For all $N\ge1$ and $k\ge1$, every zero-error protocol for RABC that uses a one-bit message and shared randomness, but no other resource, satisfies $\mathcal W\le2^{-k}$. The value $2^{-k}$ is attained.
\end{lemma}

\begin{proof} All local randomness can be absorbed into the shared random variable $\lambda$. For fixed $\lambda$, Alice's message is a function $M=f_\lambda(\bm Z)$, and Bob certifies block $i$ after message $m$ if and only if $c_\lambda(m,i)=1$, for some function $c_\lambda$. The value $\mathcal W$ is the average over $\lambda$ of the value $\mathcal W_\lambda$ of the deterministic strategy $(f_\lambda,c_\lambda)$. Moreover, the protocol has zero error only if every strategy that occurs with positive probability does. It therefore suffices to show $\mathcal W_\lambda\le2^{-k}$ for a zero-error deterministic strategy, and we drop the index $\lambda$.

Every string $\bm Z\in\{0,1\}^{kN}$ occurs with probability $2^{-kN}>0$. Hence zero error [Eq.~\eqref{eq:zeroerr}] requires
that $c(m,i)=1$ only if $\bm Z_i=1^k$ for every $\bm Z$ with $f(\bm Z)=m$.

Let $m_0=f(0^{kN})$ be the message sent for the all-zero string. No block of $0^{kN}$ equals $1^k$, so $c(m_0,i)=0$ for every $i$. Let $\bar m$ be the other message and $T=\{i:c(\bar m,i)=1\}$. Bob certifies block $i$ exactly when $i\in T$ and $f(\bm Z)=\bar m$, so
\[
\mathcal W=\sum_{i=1}^N\Pr(C_i=1)=|T|\,\Pr\big(f(\bm Z)=\bar m\big).
\]

By the zero-error condition, every string with $f(\bm Z)=\bar m$ satisfies $\bm Z_i=1^k$ for all $i\in T$. The blocks are independent and uniformly distributed, so
\begin{align*}
\Pr\big(f(\bm Z)=\bar m\big)&\le\Pr\big(\bm Z_i=1^k\ \forall\, i\in T\big)=2^{-k|T|},
\end{align*}
and hence $\mathcal W\le g(|T|)$ with $g(t)\equiv t\,2^{-kt}$. We have $g(0)=0$, and for every integer $t\ge1$, $
g(t+1)/g(t)=\big(1+1/t\big)2^{-k}\le2\cdot2^{-k}\le1$. So $g$ is non-increasing on $t\ge1$, and $\mathcal W\le g(1)=2^{-k}$. The parties share a uniformly random index $\ell\in\{1,\dots,N\}$. Alice sends $M=1$ if $\bm Z_\ell=1^k$ and $M=0$ otherwise. Bob certifies block $i$ if $i=\ell$ and $M=1$. A certificate for block $i$ then requires $\bm Z_i=1^k$, so the protocol has zero error, and
\[
\mathcal W=\sum_{i=1}^N\Pr(\ell=i)\Pr(\bm Z_i=1^k)=\sum_{i=1}^N\frac1N\,2^{-k}=2^{-k}, 
\]
therby completing the proof.
\end{proof}

\end{document}